\documentclass[epjCONF,onecolumn,final]{svjour}
\usepackage{graphics}
\usepackage[varg]{txfonts} % Times fonts
\usepackage[latin1]{inputenc}
\session-title{Hot and Cold Baryonic Matter -- HCBM 2010}
\begin{document}
\title{Second order dissipative fluid dynamics from kinetic theory}
\author{B.\ Betz \inst{1} 
\and G.S.\ Denicol\inst{2} \and T.\ Koide\inst{3} 
\and E.\ Moln\'ar \inst{3,4} \and H.\ Niemi\inst{3} \and D.H.\ Rischke\inst{2,3}}
\institute{Department of Physics, Columbia University, New York, 10027, USA 
\and Institute f\"ur Theoretische Physik, Johann Wolfgang
Goethe-Universit\"at, Max-von-Laue Str. 1, D-60438, Frankfurt am Main, Germany 
\and Frankfurt Institute for Advanced Studies, Ruth-Moufang Str. 1, D-60438, 
Frankfurt am Main, Germany
\and KFKI Research Institute for Particle and Nuclear Physics,
H-1525 Budapest, P.O.Box 49, Hungary}

\abstract{
We derive the equations of second order dissipative fluid dynamics from 
the relativistic Boltzmann equation following the method of W.~Israel 
and J.~M.~Stewart \cite{Israel:1979wp}. 
We present a frame independent calculation of all first- and second-order terms and their coefficients 
using a linearised collision integral. 
Therefore, we restore all terms that were previously neglected in 
the original papers of W.~Israel and J.~M.~Stewart.
}
\maketitle

%%%
\section{Introduction}

Relativistic fluid dynamics has been applied successfully
to describe the dynamics of the hot and dense matter
created in relativistic heavy-ion collisions at the Relativistic
Heavy Ion Collider (RHIC) \cite{Huovinen:2006jp}.
It is also expected to play an important role in understanding 
of future experiments at the Large Hadron Collider (LHC) and 
at the Facility for Antiproton and Ion Research (FAIR).

On the other hand, the theoretical foundations of relativistic dissipative
fluid dynamics are not fully established yet.
Relativistic kinetic theory of dilute gases provides a framework
for relativistic fluid dynamics which can be 
derived systematically from the Boltzmann equation. 
However, this procedure is not unique and has been subject of many 
past and recent studies. 

In this paper we present a short but self-contained derivation of 
dissipative fluid dynamics from kinetic theory following the widely 
used approach by Israel and Stewart (IS) which is a generalization 
of Grad's method of moments \cite{Grad} to relativistic systems.
We show that the resulting equations contain several 
new second order terms and coefficients \cite{Betz:2008me} which were 
absent in the treatment by IS and others in the past.

%%%
Notation: 
We define the space-time coordinates choosing $c=1$, and the covariant and contravariant  
four-vectors, $x_{\mu} = g_{\mu \nu} x^{\nu} \equiv  (t,-x,-y,-z)$ and 
$x^{\mu} = g^{\mu \nu} x_{\nu} \equiv (t,x,y,z)$, 
where $g^{\mu \nu} = g_{\mu \nu} = \textrm{diag} (1,-1,-1,-1)$ 
is the metric of flat space-time.
The normalized hydrodynamic four-velocity of matter is $u^{\mu}$, such that $u^{\mu} u_{\mu} = 1$. 
The transverse projection operator $\Delta^{\mu \nu} \equiv g^{\mu \nu} - u^{\mu} u^{\nu}$, 
is used to decompose four-vectors or tensors into parts parallel and 
orthogonal to $u^{\mu}$.
The transverse projection of four-vectors is denoted by 
$A^{\langle \mu \rangle} = \Delta^{\mu}_{\nu} A^{\nu}$.
The transverse and traceless projection of second-rank tensors is defined as
$A^{\langle \mu \nu \rangle} 
\equiv \Delta^{\mu \nu}_{\alpha \beta} A^{\alpha \beta}
\equiv \left[\frac{1}{2}\left(\Delta^{\mu}_{\alpha} \Delta^{\nu}_{\beta} 
+ \Delta^{\nu}_{\alpha} \Delta^{\mu}_{\beta} \right) 
- \frac{1}{3} \Delta^{\mu \nu} \Delta_{\alpha \beta} \right]A^{\alpha \beta}$.
The gradient, $\partial_\mu \equiv \partial/\partial x^{\mu}$, 
of an arbitrary tensor can be decomposed as,
$\partial_{\alpha} A^{\mu_1 \ldots \mu_n} 
= u_{\alpha} D A^{\mu_1 \ldots \mu_n} + \nabla_{\alpha} A^{\mu_1 \ldots \mu_n}$,
where the comoving time-derivative, $D \equiv u^{\mu} \partial_{\mu}$, is
also denoted by an over-dot, $\dot{A}^{\mu} \equiv D A^{\mu}$, 
and $\nabla_{\alpha} \equiv \Delta^{\beta}_{\alpha} \partial_{\beta}$ 
is the gradient operator.
The symmetric and antisymmetric parts of a second rank tensors are denoted by
$A^{(\mu \nu)} \equiv \left(A^{\mu \nu} + A^{\nu \mu} \right)/2$
and 
$A^{[\mu \nu]} \equiv \left(A^{\mu \nu} - A^{\nu \mu} \right)/2$, respectively.
Using the above notations the relativistic generalization of the 
Cauchy-Stokes decomposition is
$\partial_{\mu} u_{\nu} = u_{\mu} \dot{u}_{\nu} 
+ \frac{1}{3} \theta \Delta_{\mu \nu} + \sigma_{\mu \nu} + \omega_{\mu \nu}$, 
where we defined the expansion scalar $\theta \equiv \nabla_{\mu} u^{\mu} $, 
the shear tensor $\sigma^{\mu \nu} \equiv \nabla^{\langle \mu} u^{\nu \rangle}$, 
and the vorticity tensor $\omega^{\mu \nu} \equiv \nabla^{[ \mu} u^{\nu ]}$.

%%%
\section{Fluid dynamics from kinetic theory}

In relativistic kinetic theory of single component gases,
particles with mass $m$ and four-momenta $p^{\mu} = (p^0, \vec{p})$
where $p^0 = \sqrt{\vec{p}^2 + m^2}$, at 
space-time coordinate $x^{\mu} = (t,\vec{x})$ are characterized by the 
invariant single particle distribution function $f = f(t,\vec{x},\vec{p})$.
Assuming that there are no external forces, the space-time evolution 
of $f$ in a dilute gas with binary collisions is given
by the relativistic Boltzmann transport equation 
\cite{deGroot},
\begin{equation}\label{BTE}
p^{\mu} \partial_{\mu} f = C \left[f \right] \, ,
\end{equation}
where $C \left[f\right]$ is the collision integral.
The explicit form of the collision integral is
\begin{equation} \label{BUU}
C[f] = \frac{1}{2} \int d\omega_1 d\omega' d\omega'_1 \, W (p,p_1,p',p'_1) 
\left[f' f'_1 \tilde{f} \tilde{f}_1  - f f_1 \tilde{f'} \tilde{f'_1} \right] \, ,
\end{equation}
where $d\omega \equiv g d^3 p/\left[p^0 (2 \pi \hbar)^3 \right]$;
$g$ is the number of internal degrees of freedom and the transition rate 
is proportional to the cross section $W(p,p_1,p',p'_1) \sim \sigma$. 
Here $\tilde{f} \equiv 1 - af$ denotes the correction due to boson and fermion statistics 
with $a=-1$ and $a=+1$, while $a=0$ for classical gases.

For later convenience let us separate the four-momenta of particles into two parts
using an arbitrary fluid dynamic flow velocity $u^{\mu}$.
Thus $p^{\mu} = E u^{\mu} + p^{\langle \mu \rangle}$, where $E = p^{\mu} u_{\mu}$ is 
the energy of the particle in the Local Rest Frame (LRF), where $u^{\mu}_{\rm LRF} = (1,0,0,0)$, 
and $p^{\langle \mu \rangle} = \Delta^{\mu \nu} p_{\nu}$ is the LRF momentum.
The macroscopic fields such as the particle four-flow and energy-momentum tensor 
are defined as the first and second moment of the single particle distribution function,
\begin{eqnarray} \label{kinetic:N_mu}
N^{\mu}(t,\vec{x}) &\equiv& \langle p^{\mu} \rangle 
= \langle E \rangle u^{\mu} + \langle p^{\langle \mu \rangle} \rangle\, , \\
\label{kinetic:T_mu_nu}
T^{\mu \nu}(t,\vec{x}) &\equiv& \langle p^{\mu} p^{\nu} \rangle 
= \langle E^2 \rangle u^{\mu} u^{\nu} 
+ \frac{1}{3} \Delta^{\mu \nu} \langle \Delta^{\alpha \beta} p_{\alpha} p_{\beta} \rangle
+ \langle E p^{\langle \mu \rangle} \rangle u^{\nu} 
+ \langle E p^{\langle \nu \rangle} \rangle u^{\mu}
+ \langle p^{\langle \mu} p^{\nu \rangle} \rangle \, ,
\end{eqnarray}
where the brackets, $\langle \ldots \rangle \equiv \int d \omega \ldots f$, 
denote the integral of $f$ in momentum space.
Similarly, we can define higher moments or fluxes, these are 
$F^{\mu_1 \ldots \mu_n} \equiv \langle p^{\mu_1} \ldots p^{\mu_n} \rangle$, 
and they obey the recurrence relation 
$F^{\mu_1 \ldots \mu_{n}} g_{\mu_{n-1} \mu_n} = m^2 F^{\mu_1 \ldots \mu_{n-2}}$,
where $p^{\mu} p_{\mu} = m^2$ was used.

The conservation of particle number or charge and
of energy and momentum in individual collisions lead to
the equations of relativistic fluid dynamics \cite{deGroot},
\begin{eqnarray}\label{cons_N_mu}
\partial_{\mu} N^{\mu}  
&\equiv& \int d\omega \, C\left[ f\right] = 0 \, ,\qquad
\label{cons_T_mu_nu}
\partial_{\mu} T^{\mu\nu} 
\equiv \int d\omega \, p^{\nu} C \left[ f\right] = 0 \, ,
\end{eqnarray}
while the divergence of higher moments leads to the balance of fluxes,
\begin{eqnarray}
\label{cons_F_mu_1_n}
\partial_{\lambda} F^{\mu_1 \ldots \mu_{n} \lambda} 
&\equiv& \int d\omega p^{\mu_1} \ldots p^{\mu_n} C \left[ f\right] 
= P^{\mu_1 \ldots \mu_n}\, ,
\end{eqnarray}
where the $P^{\mu_1 \ldots \mu_{n}}$ denotes the production term for the 
$n$-th moment of the Boltzmann equation. 
Similarly to $F^{\mu_1 \ldots \mu_n}$, the production term also satisfies the recurrence relation,
$P^{\mu_1 \ldots \mu_{n}} g_{\mu_{n-1} \mu_n} = m^2 P^{\mu_1 \ldots \mu_{n-2}}$.

The above equations are valid for an arbitrary solution $f$ of the Boltzmann equation.
However, in case the microscopic time and length scales are much shorter than the macroscopic ones, 
the system relaxes to the so-called local equilibrium state.
In local equilibrium the single particle distribution function has a specific isotropic 
form resulting from a locally maximal entropy with vanishing entropy production, 
\begin{equation}
f_0 = \left[\exp(-\alpha_0 + \beta_0 E) + a \right]^{-1} \, ,
\end{equation}
where $\beta_0 = 1/T$ is the inverse temperature and $\alpha_0=\mu/T$ is 
the ratio of chemical potential $\mu$ and temperature, both in units of $k_{B}=1$.
Using the local equilibrium distribution function in 
Eqs. (\ref{kinetic:N_mu},\ref{kinetic:T_mu_nu}) we define, 
\begin{equation} \label{equilibrium_dens}
n_0 =\langle E \rangle_0 \, , \qquad 
e_0 =\langle E^2 \rangle_0 \, , \qquad
p_0 = - \frac{1}{3} \langle \Delta^{\alpha \beta} p_{\alpha} p_{\beta} \rangle_0 \, ,
\end{equation}
where the brackets, $\langle \ldots \rangle_0 \equiv \int d \omega \ldots f_0$, 
and $n_0 = N^{\mu}_0 u_{\mu}$ is the particle number density, 
$e_0 = T^{\mu \nu}_0 u_{\mu} u_{\nu}$ is the energy density 
and $p_0 = -T^{\mu \nu}_0 \Delta_{\mu \nu}/3$ is the isotropic pressure,
in equilibrium. 
Here we introduced the particle four-current
$N^{\mu}(f_0) \equiv N^{\mu}_0$ and energy-momentum tensor
$T^{\mu \nu}(f_0) \equiv T^{\mu \nu}_0$.
Thus the local equilibrium approximation defines a perfect fluid with only five unknown fields
$u^{\mu}$, $n_0$ and $e_0$, such that
\begin{eqnarray}
N^{\mu}_0 &\equiv& n_0 u^{\mu} \, ,\\ 
T^{\mu \nu}_0 &\equiv& e_0 u^{\mu} u^{\nu} - p_0 \Delta^{\mu \nu}\, ,
\end{eqnarray}
where the pressure $p_0$ is defined through an Equation of State (EOS).
The temperature, chemical potential and entropy are also defined 
by the laws of equilibrium thermodynamics.

If the system is out-of equilibrium the distribution function changes. 
Let us denote this difference by, $\delta f = f - f_0$, therefore the averages in 
Eqs. (\ref{kinetic:N_mu},\ref{kinetic:T_mu_nu}) define 14 new fields on the l.h.s.,
\begin{eqnarray}
n &\equiv& \langle E \rangle 
= \langle E \rangle_0 + \langle E \rangle_\delta = n_0 + \delta n  \, , \\
e &\equiv& \langle E^2 \rangle 
= \langle E^2 \rangle_0 + \langle E^2 \rangle_\delta = e_0 + \delta e \, , \\
p &\equiv& - \frac{1}{3} \langle \Delta^{\alpha \beta} p_{\alpha} p_{\beta} \rangle 
= - \frac{1}{3} \langle \Delta^{\alpha \beta} p_{\alpha} p_{\beta} \rangle_0 
- \frac{1}{3} \langle \Delta^{\alpha \beta} p_{\alpha} p_{\beta} \rangle_\delta 
= p_0 + \Pi \, , \\
V^{\mu} &\equiv& \langle p^{\langle \mu \rangle}  \rangle 
= \langle p^{\langle \mu \rangle}  \rangle_\delta \, ,\\
W^{\mu} &\equiv& \langle E p^{\langle \mu \rangle} \rangle 
= \langle E p^{\langle \mu \rangle}  \rangle_\delta \, ,\\
\pi^{\mu \nu} &\equiv& \langle p^{\langle \mu} p^{\nu \rangle} \rangle 
= \langle p^{\langle \mu} p^{\nu \rangle} \rangle_\delta \, ,
\end{eqnarray}
where $\langle \ldots \rangle_\delta \equiv \int d \omega \ldots \delta f$, 
and $n = N^{\mu} u_{\mu} = n_0 + \delta n$ is the particle number density, 
$e = T^{\mu \nu} u_{\mu} u_{\nu} = e_0 + \delta e$ is the energy density 
and $p = p_0 + \Pi = -T^{\mu \nu} \Delta_{\mu \nu}/3$ is the 
isotropic pressure decomposed into equilibrium pressure
and bulk viscous pressure parts.
Here, $V^{\mu} = \Delta^{\mu \alpha} N_{\alpha}$ 
is the particle diffusion current with $V^{\mu}u_{\mu} = 0$, 
$W^{\mu} = \Delta^{\mu \alpha} T_{\alpha \beta} u^{\beta}$ 
is the energy-momentum diffusion current with $W^{\mu}u_{\mu}=0$, 
while the orthogonal and traceless part of $T^{\mu \nu}$ defines the stress tensor,
$\pi^{\mu \nu} = T^{\langle \mu \nu \rangle}$, 
where $\pi^{\mu \nu} u_{\mu} = 0$ and $\pi^{\mu \nu} g_{\mu \nu}=0$.
Now, we can write the fundamental fluid dynamical quantities as
\begin{eqnarray} \label{N_mu}
N^{\mu} &\equiv& N^{\mu}_0 + \delta N^{\mu} 
= (n_0 + \delta n) u^{\mu} + V^{\mu} \, , \\ \label{T_munu}
T^{\mu \nu} &\equiv& T^{\mu \nu}_0 + \delta T^{\mu \nu} 
= (e_0 + \delta e) u^{\mu} u^{\nu} - (p_0 + \Pi) \Delta^{\mu \nu}
+  2 W^{(\mu}u^{\nu)} +  \pi^{\mu \nu} \, .
\end{eqnarray} 
At this point the flow of matter $u^{\mu}$ is usually specified using the 
definition of Eckart \cite{Eckart:1940te} or of Landau and Lifshitz \cite{Landau_book}.
Eckart chose the flow of conserved particles (if there are any) to define
$u^{\mu} = N^{\mu}/\!\!\sqrt{N^{\mu} N_{\mu}}$, 
while Landau and Lifshitz chose the flow of energy-momentum
$u^{\mu} = T^{\mu \nu} u_{\nu}/\!\!\sqrt{T^{\mu \alpha} u_\alpha T_{\mu \beta} u^{\beta}}$.
This implies that either the particle diffusion current or the energy-momentum diffusion 
current vanishes, $V^{\mu}=0$ or $W^{\mu}=0$. 
These physically different choices are related to each other and define the heat-flow, 
\begin{equation} \label{heat-flow}
q^{\mu} \equiv W^{\mu} - V^{\mu} (e + p)/n 
\simeq W^{\mu} - V^{\mu} (e_0 + p_0)/n_0 + O_2 \, .
\end{equation}
In the following we will not restrict the calculations by fixing a frame of reference.

%%%
\section{The method of moments}

One of the methods successfully used to derive the equations of dissipative 
fluid dynamics was introduced by H.~Grad \cite{Grad,Kemer_1990}    and later generalized to relativistic 
systems by Israel and Stewart \cite{Israel:1979wp,Stewart:1972hg,IS_1976,Stewart:1977}.
This method uses the conservation laws together with the balance equation for the third moment 
and its production term to obtain the equations of motion for the 14-fields.

Let us specify the distribution function as $f(y) = (e^{-y} + a)^{-1}$, where 
the equilibrium distribution function is recovered when $y = y_0 \equiv \alpha_0 - \beta_0 E$.
If the system is not too far away from local equilibrium the 
argument of the equilibrium distribution function changes by 
$\delta y = y - y_0 \ll 1$, therefore
\begin{equation} \label{f_expanded}
f(y) = f(y_0) + f'(y_0) \delta y + f''(y_0) \delta y^2/2 + \ldots + O_n\, ,
\end{equation}
where $f'(y_0) = f_0 \tilde{f}_0$.
The method of IS specifies the deviation from equilibrium in terms of a polynomial expansion in 
the momentum four-vector $p^{\mu}$ up to quadratic order,
\begin{equation} \label{q_expanded}
\delta y \equiv \alpha - p^{\mu} \beta_{\mu} + p^{\mu} p^{\nu} w_{\mu \nu}
= \alpha - \beta E - p^{\mu} v_{\mu} + E^2 w - \frac{w}{3} \Delta^{\mu \nu} p_{\mu} p_{\nu}
+ 2 E w^{\mu}p_{\mu} + w^{\langle \mu \nu \rangle} p_{\mu} p_{\nu}\, ,
\end{equation}
where $\beta = \beta^{\mu} u_{\mu}$ and 
$w = w^{\mu \nu} u_{\mu} u_{\nu}= -w^{\mu \nu} \Delta_{\mu \nu}$ are scalars, 
$v^{\mu} = \beta^{\langle \mu \rangle}$ and $w^{\mu} = w^{\langle \mu \rangle \beta} u_{\beta}$ 
are four-vectors orthogonal to $u^{\mu}$, and $ w^{\langle \mu \nu \rangle}$ 
is the traceless and orthogonal part of $w^{\mu \nu}$.

Substituting the previous two equations into Eqs. (\ref{kinetic:N_mu},\ref{kinetic:T_mu_nu})
we get
\begin{eqnarray} \label{N_mu:Grad} 
N^{\mu} &=& \left[I_{10} + \alpha J_{10}
- \beta J_{20} + w (J_{30} - J_{31}) \right] u^{\mu} 
- J_{21} v^{\mu} + 2 J_{31} w^{\mu} \, , \\ \nonumber
\label{T_munu:Grad}
T^{\mu \nu} &=& \left[I_{20} + \alpha J_{20}
- \beta J_{30} + w(J_{40} - J_{41})\right] u^{\mu} u^{\nu}
- \left[-I_{21} - \alpha J_{21}
+ \beta J_{31} - w\left(J_{41} - \frac{5}{3} J_{42} \right) \right] \Delta^{\mu \nu} \\
&-& 2\left(J_{31} v^{(\mu} - 2J_{41} w^{(\mu}\right) u^{\nu)}
+ 2 J_{42} w^{\langle \mu \nu \rangle}  \, .
\end{eqnarray}
Here we introduced the auxiliary thermodynamic integrals, $I_{nq}(\alpha_0,\beta_0)$ and 
$J_{nq}(\alpha_0,\beta_0)$,
\begin{equation}
I_{nq} \equiv \frac{1}{(2q +1)!!} \int d\omega \, E^{n-2q} 
\left( p^{\mu} p^{\nu} \Delta_{\mu \nu} \right)^{q} f_0 \, , \quad
J_{nq} \equiv \frac{1}{(2q +1)!!} \int d\omega \, E^{n-2q} 
\left( p^{\mu} p^{\nu} \Delta_{\mu \nu} \right)^{q} f_0 \tilde{f}_0 \, ,
\end{equation}
where $n$ and $q \leq \left[n/2 \right]$ are integers and $(2q+1)!!$ denotes the double factorial. 
For a classical gas $J_{nq}(a=0) = I_{nq}$.
These two integrals are related by 
$\beta_0 J_{nq} = -I_{n-1,q-1} + (n-2q) I_{n-1,q}$,
$d I_{nq} = J_{nq} d\alpha_0 - J_{n+1,q} d\beta_0$, 
and $\beta_0 d J_{nk} = \left[ -J_{n-1,q-1} + (n-2q)J_{n-1,q} \right] d\alpha_0 
+ \left[ J_{n,q-1} - (n-2q + 1)J_{nq} \right] d\beta_0$ where $d$ stands for both the proper time and
spatial derivatives \cite{Muronga:2006zw}.

Furthermore, comparing Eqs. (\ref{N_mu}, \ref{T_munu}) 
with Eqs. (\ref{N_mu:Grad}, \ref{T_munu:Grad}) we can express the fluid dynamical fields in terms 
of the thermodynamic integrals and the parameters of the distribution function, 
\begin{eqnarray}
n_0 &=& I_{10} \, , \qquad e_0 = I_{20} \, , \qquad p_0 = -I_{21} \, , \label{dn} \\ 
\delta n &=& \alpha J_{10} - \beta J_{20} + w(J_{30} - J_{31}) \, , \qquad
\delta e = \alpha J_{20} - \beta J_{30} + w(J_{40} - J_{41}) \, , \label{dp}\\
\Pi &=&  - \alpha J_{21} + \beta J_{31} - w\left(J_{41} - 
\frac{5}{3} J_{42}\right) \, , \\ 
V^{\mu} &=& - J_{21} v^{\mu} + 2 J_{31} w^{\mu} \, , \qquad 
W^{\mu} = - J_{31} v^{\mu} + 2J_{41} w^{\mu} \, , \\
\pi^{\mu \nu} &=& 2 J_{42} w^{\langle \mu \nu \rangle} \, .
\end{eqnarray}
The above relations can be inverted to extract the parameters of the non-equilibrium 
distribution function in terms of the $14$ fluid dynamical fields,
$n=n_0 + \delta n, e=e_0 + \delta e, V^{\mu}, W^{\mu}, \pi^{\mu \nu}$ and $p=p_0 + \Pi$.
Without loss of generality the scalar quantities can be inferred from Eqs. (\ref{dn},\ref{dp}), 
however the current method does not provide enough independent equations for closure.
Therefore, one usually employs a so-called matching or fitting condition where 
the non-equilibrium particle and energy densities are assumed to be unchanged 
at least up to first order in deviations from equilibrium. 
Hence, following IS we fix $\delta n = \delta e = 0$.
This also means that the EOS is given by, $p_0 = p_0(e_0,n_0)$, 
and the temperature and chemical potential are fixed by the equilibrium state.
Alternatively one could use different matching conditions \cite{Tsumura:2007ji} or even
different thermodynamical theories, see for example Refs. \cite{Gariel:1994nw,Hayward:1998hb,Van:2007pw}.
Thus from the above equations and constraints we get
\begin{eqnarray} \label{abw}
\alpha &=& \mathcal{A}_{\Pi \alpha} \Pi \, ,\qquad
\beta = \mathcal{A}_{\Pi \beta} \Pi\, ,\qquad 
w = - \mathcal{A}_{\Pi w} \Pi \, , \\  \label{VW}
v^{\mu} &=& - \mathcal{B}_{V v} V^{\mu} + \mathcal{B}_{W v} W^{\mu} \, , \qquad
w^{\mu} =  - \mathcal{B}_{V w} V^{\mu} + \mathcal{B}_{W w} W^{\mu} \, , \\ \label{pi_w}
w^{\langle \mu \nu \rangle}  &=& \mathcal{C}_{\pi w} \pi^{\mu \nu} \, ,
\end{eqnarray}
where we introduced
\begin{eqnarray} 
\mathcal{A}_{\Pi \alpha} 
&=& \mathcal{A}_{\Pi w} \left[m^2 - 4 \left( J_{30}J_{31} - J_{20}J_{41}\right)/D_{20} \right]\, , \qquad
\mathcal{A}_{\Pi \beta} 
= \mathcal{A}_{\Pi w} \left[4 \left(J_{10}J_{41} - J_{20}J_{31}\right)/D_{20} \right]\, , \qquad \\
\mathcal{A}_{\Pi w} &=& \frac{- 3 D_{20}  \Pi}{4\left[3J_{21} (J_{30}J_{31} - J_{20}J_{41})
+ 3J_{31}(J_{10} J_{41} - J_{20} J_{31}) - 5 J_{42} D_{20} \right]} \, , \\ 
\mathcal{B}_{V v} &=& J_{41}/D_{31} \, , \qquad \mathcal{B}_{W v} = J_{31}/D_{31} \, , \qquad
\mathcal{B}_{V w} = J_{31}/(2 D_{31}) \, , \qquad \mathcal{B}_{W w} = J_{21}/(2 D_{31}) \, , \\ 
\mathcal{C}_{\pi w} &=& \left(2 J_{42} \right)^{-1} \, ,
\end{eqnarray}
and $D_{nq} = J_{n-1,q} J_{n+1,q} - J^2_{nq}$. 

Here we note that using the definition of heat-flow together with the definition 
of enthalpy per particle
$h \equiv (e_0 + p_0)/n_0 = J_{31}/J_{21}$, one can equivalently choose
$v^{\mu} = - \mathcal{B}'_{V v} V^{\mu} + \mathcal{B}'_{q v} q^{\mu}$ 
and $w^{\mu} = \mathcal{B}'_{q w} q^{\mu}$, where 
$\mathcal{B}'_{V v} \equiv (J_{21})^{-1}$, 
$\mathcal{B}'_{q v} \equiv J_{31}/D_{31} = \mathcal{B}_{Wv}$ and  
$\mathcal{B}'_{q w} \equiv J_{21}/(2 D_{31}) = \mathcal{B}_{Ww}$,
however we prefer to write the equations in terms of $V^{\mu}$ and $W^{\mu}$.

%%%
\subsection{The balance equations from the third moment}

The equations of motion for the nine dissipative fields can be calculated 
from the third or higher moment of the Boltzmann equation.
Therefore using Eq. (\ref{cons_F_mu_1_n}) for $n \geq 2$, the equations for 
bulk viscous pressure, heat and diffusion currents, and stress tensor can be formally written as
\begin{eqnarray} \label{scalar_eq}
 u_{\mu_1} \ldots u_{\mu_n} \partial_\lambda F^{\mu_1 \ldots \mu_n \lambda} 
&=& u_{\mu_1} \ldots u_{\mu_n} P^{\mu_1 \ldots \mu_n} \, ,\\ \label{vector_eq}
\Delta^{\alpha}_{\mu_1} u_{\mu_2} \ldots u_{\mu_n} \partial_\lambda F^{\mu_1 \ldots \mu_n \lambda}
&=& \Delta^{\alpha}_{\mu_1} u_{\mu_2} \ldots u_{\mu_n}  P^{\mu_1 \ldots \mu_n} \, , \\ \label{tensor_eq}
\Delta^{\alpha \beta}_{\mu_1 \mu_2} u_{\mu_3} \ldots u_{\mu_n} 
\partial_\lambda F^{\mu_1 \ldots \mu_n \lambda}
&=& \Delta^{\alpha \beta}_{\mu_1 \mu_2}  u_{\mu_3} \ldots u_{\mu_n} P^{\mu_1 \ldots \mu_n} \, .
\end{eqnarray}
Here we follow the method of IS and augment the equations 
of fluid dynamics Eqs.\ (\ref{cons_N_mu}) which are 
Eqs.\ (\ref{cons_F_mu_1_n}) for $n=0$ and $n=1$ with the next equation for $n=2$ that is 
$\partial_\lambda F^{\mu \nu \lambda} = P^{\mu \nu}$. 
Using the 14-moment approximation, $F^{\mu \nu \lambda}$ and 
its production term $P^{\mu \nu}$ become functions of the nine dissipative fields.
This means that one discards irreducible tensors
of rank higher than two, which usually do not appear in fluid dynamics.
Of course one could obtain the equations of motion from 
higher moments of the Boltzmann equation,
however, we note that the resulting equations of motion
would be formally identical but yield different transport coefficients
as shown in Ref. \cite{Denicol:2010xn}.

In the following we explicitly calculate the equations of motion for the dissipative fields 
as done by Israel and Stewart, but here we restore all the terms that they neglected.
Hence, using Eqs. (\ref{abw},\ref{VW},\ref{pi_w}) the third moment leads to
\begin{eqnarray} \nonumber
F^{\mu \nu \lambda}
\equiv \langle p^{\mu} p^{\nu} p^{\lambda} \rangle &=& \left(I_{30} + \psi_4 \Pi \right) u^\mu u^\nu u^\lambda
+ 3 \left(I_{31} - \psi_4 \Pi/3 \right) u^{(\mu} \Delta^{\nu \lambda)}
+ 3 \psi^W_1 W^{(\mu} u^{\nu} u^{\lambda)} + 3 \psi^V_1 V^{(\mu} u^{\nu} u^{\lambda)} \\ 
&+& 3 \psi^W_2 W^{(\mu} \Delta^{\nu \lambda)} + 3 \psi^V_2 V^{(\mu} \Delta^{\nu \lambda)} 
+ 3 \psi_3 \pi^{(\mu \nu} u^{\lambda)}\, ,
\end{eqnarray}
where we introduced the following variables
\begin{eqnarray}
\psi_4 &=& -\frac{J_{30}(J_{30}J_{31} - J_{20} J_{41})
+ J_{40}(J_{10}J_{41} - J_{20} J_{31}) - J_{51}D_{20}}
{J_{21}(J_{30} J_{31} - J_{20}J_{41}) +  J_{31}(J_{10}J_{41} - J_{20}J_{31})- 5 J_{42}D_{20}/3}\, ,\\
\psi^W_1 &=& \left(J_{21}J_{51} - J_{31}J_{41}\right)/D_{31} \, , \qquad \psi^W_2 = - \psi^W_1/5 \, , \\
\psi^V_1 &=& -D_{41}/D_{31} \, , \qquad \psi^V_2 = m^2/5 - \psi^V_1/5 \, , \\
\psi_3 &=& J_{52}/J_{42}\, .
\end{eqnarray}
The equations of motion follow from the different projections of the balance 
equation of the third moment.
Thus Eq. (\ref{scalar_eq}) leads the equation for the bulk viscous pressure $\Pi$,
\begin{eqnarray} \label{bulk_eq_motion} \nonumber
u_{\mu} u_{\nu} P^{\mu \nu}  &=& \dot{I}_{30} + \psi_4 \dot{\Pi} + \dot{\psi_4} \Pi 
+ \left(I_{30} - 2I_{31} + \frac{5}{3} \psi_4 \Pi \right) \theta \\ 
&+& \partial_{\mu} (\psi^W_1 W^{\mu}) - 2\psi^W_1 W_{\mu} \dot{u}^{\mu}
+ \partial_{\mu} (\psi^V_1 V^{\mu}) - 2\psi^V_1 V_{\mu} \dot{u}^{\mu}
- 2\psi_3 \pi^{\mu \nu} \partial_{\mu} u_{\nu} \, .
\end{eqnarray}
The vector equation Eq. (\ref{vector_eq}) for $W^{\mu}$ and $V^{\mu}$ is
\begin{eqnarray} \label{heat_eq_motion} \nonumber 
u_{\nu} \Delta^{\mu}_{\alpha} P^{\nu \alpha} 
&=& \left(I_{30} - 2I_{31} + \frac{5}{3} \psi_4 \Pi \right)\dot{u}^{\mu}
+ W^{\mu} \left[\dot{\psi}^W_1 + (\psi^W_1 - \psi^W_2)\theta \right]
+ V^{\mu} \left[\dot{\psi}^V_1 + (\psi^V_1 - \psi^V_2)\theta \right] \\ \nonumber
&+& \Delta^{\mu}_{\alpha} \left[\partial^{\alpha} \left(I_{31} - \frac{1}{3}\psi_4 \Pi \right) 
- \psi^W_2 W^{\nu} \partial^{\alpha} u_{\nu} + \psi^W_1 \dot{W}^\alpha 
- \psi^V_2 V^{\nu} \partial^{\alpha} u_{\nu} + \psi^V_1 \dot{V}^\alpha \right] \\ 
&+& (\psi^W_1 - \psi^W_2) W^{\nu} \partial_{\nu} u^{\mu} 
+ (\psi^V_1 - \psi^V_2) V^{\nu} \partial_{\nu} u^{\mu} 
+ (\partial_{\nu} \psi_3 - \psi_3 \dot{u}_{\nu})\pi^{\mu \nu} 
+ \psi_3 \Delta^{\mu}_{\alpha}\partial_{\nu} \pi^{\alpha \nu} \, .
\end{eqnarray}
The equation for the stress tensor $\pi^{\mu \nu}$ can be calculated from Eq. (\ref{tensor_eq}),  
\begin{eqnarray} \label{shear_eq_motion} \nonumber
P^{\langle \mu \nu \rangle}
&=& 2 \left( I_{31} - \frac{1}{3} \psi_4 \Pi \right) \partial^{\langle \mu} u^{\nu \rangle}
+ 2(\psi^W_1 - \psi^W_2) \dot{u}^{\langle \mu} W^{\nu \rangle} 
+ 2(\psi^V_1 - \psi^V_2) \dot{u}^{\langle \mu} V^{\nu \rangle} 
+ 2\left( W^{\langle \mu} \partial^{\nu \rangle} \psi^W_2
+ \psi^W_2 \partial^{\langle \nu} W^{\mu \rangle}\right) \\
&+& 2\left( V^{\langle \mu} \partial^{\nu \rangle} \psi^V_2
+ \psi^V_2 \partial^{\langle \nu} V^{\mu \rangle}\right) 
+ \dot{\psi}_3 \pi^{\mu \nu} 
+ \psi_3 \left(\dot{\pi}^{\langle \mu \nu \rangle} + \pi^{\mu \nu} \theta \right) 
+ 2 \psi_3 \pi^{\lambda \langle \mu} \partial_{\lambda} u^{\nu \rangle} \, .
\end{eqnarray}

Next we evaluate the production term using the r.h.s. of the Boltzmann equation.
Using the distribution function from Eq. (\ref{f_expanded}) in Eq. (\ref{BUU}),
the linearized production term is given by
\begin{eqnarray} \nonumber \label{lin_coll_int}
P^{\mu \nu}&\equiv&\frac{1}{2} \int d \omega d\omega_1  d\omega' d\omega'_1 \, W(p,p_1,p',p'_1) \, 
f'_{0} f'_{1,0} f_{0} f_{1,0} \, e^{-(y'_{0} + y'_{1,0} + y_{0} + y_{1,0})/2} \\ 
&\times& p^{\mu} p^{\nu} \left[p'^{\alpha} p'^{\beta} + p'^{\alpha}_1 p'^{\beta}_1 
- p^{\alpha} p^{\beta} - p^{\alpha}_1 p^{\beta}_1 \right] w_{\alpha \beta} 
= C^{\mu \nu \alpha \beta} w_{\alpha \beta} \, ,
\end{eqnarray}
where the collision tensor, $C^{\mu \nu \alpha \beta} = C^{(\mu \nu) (\alpha \beta)}$,
is symmetric upon the interchange of two incoming or outgoing particles.
Furthermore, the collision tensor is traceless 
$C^{\mu \nu \alpha \beta} g_{\alpha \beta} = C^{\mu \nu \alpha \beta} g_{\mu \nu} = 0$, 
and obeys time-reversal symmetry, 
$C^{\mu \nu \alpha \beta} = C^{\alpha \beta \mu \nu }$.
Using these properties we decompose the collision tensor as, 
\begin{equation}
C^{\mu \nu \alpha \beta} = \frac{A_0}{3} \left[3 u^{\mu} u^{\nu} u^{\alpha} u^{\beta} 
- \left(u^{\mu}u^{\nu}\Delta^{\alpha \beta} 
+ u^{\alpha}u^{\beta}\Delta^{\mu \nu} \right) 
+ \frac{1}{3} \Delta^{\mu \nu} \Delta^{\alpha \beta} \right] 
+ 4B_0 u^{(\mu} \Delta^{\nu) (\alpha} u^{\beta)} 
+ \frac{C_0}{5} \Delta^{\mu \langle \alpha} \Delta^{\beta\rangle \nu} \, ,
\end{equation}
where 
$A_0 = C^{\mu \nu \alpha \beta} u_{\mu} u_{\nu} u_{\alpha} u_{\beta}$,
$B_0 = C^{\mu \nu \alpha \beta} u_{(\mu} \Delta_{\nu) (\alpha} u_{\beta)}/3$,
$C_0 = C^{\mu \nu \alpha \beta} \Delta_{\mu \langle \alpha} \Delta_{\beta\rangle \nu}$.
Therefore using the above equations we easily get the l.h.s. of the balance equations,
\begin{eqnarray} \label{coll_bulk}
u_{\mu} u_{\nu} P^{\mu \nu} &\equiv& 4 A_0 w/3 
= \mathcal{C}_{\Pi} \Pi\, , \\  \label{coll_heat}
u_{\nu} \Delta^{\mu}_{\alpha} P^{\nu \alpha} &\equiv& 2 B_0 w^{\mu} 
= \mathcal{C}_{V} V^{\mu} + \mathcal{C}_{W} W^{\mu}\, , \\ \label{coll_shear}
P^{\langle \mu \nu \rangle} &\equiv& C_0 w^{\langle \mu \nu \rangle} /5=
\mathcal{C}_{\pi} \pi^{\mu \nu} \, ,
\end{eqnarray}
where $\mathcal{C}_{\Pi} = -(4 A_0 \mathcal{A}_{\Pi w}/3)$, 
$\mathcal{C}_{V} = - 2 B_0 \mathcal{B}_{Vw}$, $\mathcal{C}_{W} = 2 B_0 \mathcal{B}_{Ww}$ 
and $\mathcal{C}_{\pi} = C_0 \mathcal{C}_{\pi w}/5$.

%%%
\subsection{The relaxation equations}

The relaxation equations follow from the balance equations 
(\ref{bulk_eq_motion}, \ref{heat_eq_motion}, \ref{shear_eq_motion}) 
and the linearized collision integral, Eqs. (\ref{coll_bulk},\ref{coll_heat},\ref{coll_shear}).
Here we write the equations in a frame independent form since it is easy to re-write them 
in the Eckart frame, where $V^{\mu} = 0$ and $q^{\mu} = W^{\mu}$, 
or in the Landau and Lifshitz frame, where $W^{\mu} = 0$ and then $q^{\mu} = -h V^{\mu}$.

The relaxation equation for bulk viscosity is
\begin{eqnarray} \nonumber \label{bulk_relax}
\Pi &=&  - \zeta \theta 
- \tau_{\Pi} \dot{\Pi}
+ \tau_{\Pi W} W_{\mu} \dot{u}^{\mu}
- l_{\Pi W}\partial_{\mu} W^{\mu} 
+ \lambda_{\Pi W} W^{\mu} \nabla_{\mu} \alpha_0 \\ 
&-& \zeta \delta_0 \Pi  \theta 
+ \tau_{\Pi V} V_{\mu} \dot{u}^{\mu} 
- l_{\Pi V} \partial_{\mu} V^{\mu}  
+ \lambda_{\Pi V} V^{\mu} \nabla_{\mu} \alpha_0 
+ \lambda_{\Pi \pi} \pi^{\mu \nu} \sigma_{\mu \nu} \, ,
\end{eqnarray}
where we introduced $G_{nq} = J_{n0} J_{q0} - J_{n-1,0} J_{q+1,0}$ and
\begin{eqnarray} 
\zeta &=& \frac{-1}{\mathcal{C}_\Pi D_{20}} \left(n_0 D_{30} + (e_0 + p_0) G_{23} - \beta_0 J_{41}\right) \, , \qquad
\tau_{\Pi} = - \frac{\psi_4}{\mathcal{C}_\Pi} \, , \\ 
\beta_{\Pi} &=& \tau_{\Pi}/\zeta = \psi_4 D_{20}/\left(n_0 D_{30} + (e_0 + p_0) G_{23} - \beta_0 J_{41}\right) \, ,\\ 
\tau_{\Pi W} &=& \frac{-1}{\mathcal{C}_\Pi}\left(2\psi^W_1 + \beta_0 \frac{\partial \psi^W_1}{\partial \beta_0} 
+ \frac{G_{23}}{D_{20}}\right) \, , \qquad
\tau_{\Pi V} = \frac{-1}{\mathcal{C}_\Pi}\left(2\psi^V_1 + \beta \frac{\partial \psi^V_1}{\partial \beta_0} \right) \, ,\\
l_{\Pi W} &=& \frac{-1}{\mathcal{C}_\Pi}\left(\psi^W_1 + \frac{G_{23}}{D_{20}} \right) \, ,\qquad
l_{\Pi V} = \frac{-1}{\mathcal{C}_\Pi}\left(\psi^V_1 + \frac{D_{30}}{D_{20}}\right) \, , \\
\lambda_{\Pi W} &=& \frac{1}{\mathcal{C}_\Pi}\left(\frac{\partial \psi^W_1}{\partial \alpha_0} 
+ h^{-1}\frac{\partial \psi^W_1}{\partial \beta_0} \right) \, ,\qquad
\lambda_{\Pi V} = \frac{1}{\mathcal{C}_\Pi}\left(\frac{\partial \psi^V_1}{\partial \alpha_0} 
+ h^{-1} \frac{\partial \psi^V_1}{\partial \beta_0} \right) \, , \\
\lambda_{\Pi \pi} &=& \frac{-1}{\mathcal{C}_\Pi} \left(2\psi_3 + \frac{G_{23}}{D_{20}}\right) \, , \qquad
\delta_0 = \beta_\Pi \left(\frac{\dot{\psi}_4}{\psi_4 \theta} + \frac{5}{3} + \frac{G_{23}}{\psi_4 D_{20}} \right) \, .
\end{eqnarray}
The relaxation equation for the flow of energy-momentum and conserved charge is
\begin{eqnarray} \label{heat_relax} \nonumber
W^{\mu} - h V^{\mu} 
&=& -\kappa \frac{n}{\beta^2_0 (e+p)}\nabla^{\mu} \alpha_0
- \tau_W \dot{W}^{\langle \mu \rangle} 
+ h \tau_V \dot{V}^{\langle \mu \rangle}
+ \tau_{W} W_{\nu} \omega^{\mu \nu}
- h\tau_{V} V_{\nu} \omega^{\mu \nu} \\ \nonumber
&-& \tau_{q\Pi} \Pi \dot{u}^{\mu} 
- \tau_{q\pi} \pi^{\mu \nu} \dot{u}_{\nu} 
+  l_{q\Pi} \nabla^{\mu} \Pi 
- l_{q\pi} \Delta^{\mu}_{\alpha}\partial_{\nu} \pi^{\alpha \nu} 
+ \lambda_{q \Pi}\Pi \nabla^{\mu} \alpha_0
+ \lambda_{q\pi} \pi^{\mu \nu} \nabla_{\nu} \alpha_0  \\
&-& \frac{\kappa}{\beta_0} \delta_{1W} W^{\mu} \theta 
+ h \frac{\kappa}{\beta_0} \delta_{1V} V^{\mu} \theta 
- \lambda_{WW} W_{\nu} \sigma^{\mu \nu}
+ h \lambda_{VV} V_{\nu} \sigma^{\mu \nu} \, ,
\end{eqnarray}
where we used that $\mathcal{C}_V = -h \mathcal{C}_W$ and defined, 
\begin{eqnarray}
\kappa &=& \frac{h \beta^2_0}{\mathcal{C}_W} \frac{D_{31}}{J_{31}} \, ,\qquad 
\tau_W = -\frac{1}{\mathcal{C}_W} \left(\psi^W_1 + \frac{\beta_0 J_{41}}{e_0 + p_0}\right) \, , \qquad 
\tau_V = -\frac{\psi^V_1}{\mathcal{C}_V} = \frac{\psi^V_1}{h\mathcal{C}_W}\, , \\
\beta_W &=& \beta_0 \tau_{W}/\kappa
= -\left(\psi^W_1 + \frac{\beta_0 J_{41}}{e_0 + p_0}\right)\frac{J_{31}}{h \beta_0 D_{31}} \, , \qquad 
\beta_V = \beta_0 \tau_{V}/\kappa = \frac{\psi^V_1 J_{31}}{h^2 \beta_0 D_{31}}\, , \\
\tau_{q\Pi} &=& \frac{-1}{\mathcal{C}_W} \left(\frac{5}{3}\psi_4 
+ \frac{\beta_0}{3} \frac{\partial \psi_4}{\partial \beta_0} + \frac{\beta_0 J_{41}}{e_0 + p_0}\right) \, , \qquad 
\tau_{q\pi} = \frac{1}{\mathcal{C}_W} \left(\psi_3 
+ \beta_0 \frac{\partial \psi_3}{\partial \beta_0} \right) \, ,\\
l_{q\Pi} &=& \frac{-1}{\mathcal{C}_W}\left(\frac{\psi_4}{3} + \frac{\beta_0 J_{41}}{e_0 + p_0}\right) \, , \qquad 
l_{q\pi} = \frac{-1}{\mathcal{C}_W} \left(\psi_3 + \frac{\beta_0 J_{41}}{e_0 + p_0}\right)\, , \\
\lambda_{q\Pi} &=& \frac{-1}{3 \mathcal{C}_W} \left(\frac{\partial \psi_4}{\partial \alpha_0} 
+ h^{-1} \frac{\partial \psi_4}{\partial \beta_0} \right) \, , \qquad
\lambda_{q\pi} = \frac{1}{\mathcal{C}_W} \left(\frac{\partial \psi_3}{\partial \alpha_0} 
+ h^{-1} \frac{\partial \psi_3}{\partial \beta_0} \right) \, ,\\
\lambda_{WW} &=& \frac{-1}{\mathcal{C}_W}\left(\frac{7 \psi^W_1}{5} + \frac{\beta_0 J_{41}}{e_0 + p_0}\right) \, ,\qquad
\lambda_{VV} = \frac{-1}{\mathcal{C}_V} \left(\frac{7\psi^V_1}{5} - \frac{2 m^2}{5} \right)\, , \qquad  \\
\delta_{1W} &=& \beta_{W} \left(\frac{\dot{\psi}^W_1}{\psi^W_1 \theta} + \frac{5}{3} 
+ \frac{4\beta_0 J_{41}}{3(e_0 + p_0)\psi^W_1}\right) \, , \qquad
\delta_{1V} = \beta_{V} \left(\frac{\dot{\psi}^V_1}{\psi^V_1 \theta} + \frac{5}{3} - \frac{m^2}{3}\right) \, .
\end{eqnarray}
The equation of motion  for the shear stress tensor is
\begin{eqnarray} \label{shear_relax} \nonumber
\pi^{\mu \nu}
&=& 2\eta \sigma^{\mu \nu} - \tau_{\pi} \dot{\pi}^{\langle \mu \nu \rangle} 
+ 2 \lambda_{\pi \Pi} \Pi \sigma^{\mu \nu} + 2 \tau_{\pi W} W^{\langle \mu} \dot{u}^{\nu \rangle} 
+ 2 \tau_{\pi V}  V^{\langle \mu} \dot{u}^{\nu \rangle} 
+ 2 l_{\pi W} \nabla^{\langle \mu} W^{\nu \rangle}  
+ 2 l_{\pi V} \nabla^{\langle \mu} V^{\nu \rangle} \\ 
&-& 2 \lambda_{\pi W} W^{\langle \mu} \nabla^{\nu \rangle} \alpha_0 
- 2 \lambda_{\pi V} V^{\langle \mu} \nabla^{\nu \rangle} \alpha_0 
- 2 \eta \delta_2 \pi^{\mu \nu} \theta
- 2 \tau_{\pi} \pi^{\langle \mu}_{\alpha} \sigma^{\nu \rangle \alpha} 
+ 2 \tau_{\pi} \pi^{\langle \mu}_{\alpha} \omega^{\nu \rangle \alpha} \, ,
\end{eqnarray}
where 
\begin{eqnarray}
\eta &=& \frac{I_{31}}{\mathcal{C}_{\pi}} \, , \qquad 
\tau_\pi = - \frac{\psi_3}{\mathcal{C}_{\pi}} \, , \qquad
\qquad \beta_{\pi} = \tau_{\pi}/(2\eta) = -\psi_3/(2 I_{31}) \, ,\\
\lambda_{\pi \Pi} &=& -\frac{\psi_4}{3 \mathcal{C}_{\pi}} \, , \qquad  
\lambda_{\pi W} = \frac{-1}{\mathcal{C}_{\pi}} 
\left( \frac{\partial \psi^W_2}{\partial \alpha_0} 
+ h^{-1} \frac{\partial \psi^W_2}{\partial \beta_0} \right) \, ,\qquad
\lambda_{\pi V} = \frac{-1}{\mathcal{C}_{\pi}} 
\left( \frac{\partial \psi^V_2}{\partial \alpha_0} 
+ h^{-1} \frac{\partial \psi^V_2}{\partial \beta_0} \right) \, ,\\
\tau_{\pi W} &=& \frac{1}{\mathcal{C}_{\pi}} \left(\frac{6}{5}\psi^W_1
- \beta_0 \frac{\partial \psi^W_2}{\partial \beta_0}  \right)\, , \qquad 
\tau_{\pi V} = \frac{1}{\mathcal{C}_{\pi}} \left(\frac{m^2}{5} + \frac{6}{5}\psi^V_1
- \beta_0 \frac{\partial \psi^V_2}{\partial \beta_0} \right)\, , \\ 
l_{\pi W} &=& \frac{\psi^W_2}{\mathcal{C}_{\pi}} \, , \qquad
l_{\pi V} = \frac{\psi^V_2}{\mathcal{C}_{\pi}} \, , \qquad
\delta_2 = \beta_{\pi} \left(\frac{\dot{\psi}_3}{\psi_3 \theta} + \frac{5}{3}\right) \, .
\end{eqnarray}
Note that we made extensive use of the following identities,
$\nabla^{\mu} \beta_0 = h^{-1} \nabla^{\mu} \alpha_0 - \beta_0 \dot{u}^{\mu}$ 
and $\nabla^{\mu} \psi = \left(\frac{\partial \psi}{\partial \alpha_0} 
+ h^{-1} \frac{\partial \psi}{\partial \beta_0} \right)\nabla^{\mu} \alpha_0
- \beta_0 \frac{\partial \psi}{\partial \beta_0} \dot{u}^{\mu}$.

The above relaxation equations are usually written and solved in a form which is given 
dividing the relaxation equations by their respective relaxation times.
This leads to coefficients which do not depend on the cross section.
Here we list them in the ultrarelativistic limit, where $m/T \rightarrow 0$, $e=3p$ and $\Pi=0$, 
in both the Eckart and Landau and Lifshitz frames,
\begin{eqnarray}
\beta_W = \beta_V &\rightarrow& \frac{5 n_0}{4 \beta_0 p^2_0}\, , \quad 
\beta^{\lambda}_{W W} = \frac{\lambda_{W W}}{\tau_{W}} \rightarrow \frac{9}{5} \, , \quad
\beta^{\lambda}_{V V} = \frac{\lambda_{V V}}{\tau_{V}} \rightarrow \frac{7}{5} \, , \quad \\ 
\beta^{\lambda}_{W \pi } &=& \frac{\lambda_{q \pi}}{\tau_{W}} \rightarrow \frac{3 n_0}{10 \beta_0 p_0} \, , \quad
\beta^{\lambda}_{V \pi } = \frac{\lambda_{q \pi}}{\tau_{V}} \rightarrow -h \frac{3 n_0}{40 p_0} \, , \quad
\delta_{1W} = \frac{2}{3} \beta_{W} \, , \quad \delta_{1V} = \beta_{V} \\
\beta^{\tau}_{W \pi} &=& \frac{\tau_{q \pi}}{\tau_{W}} \rightarrow 0 \, , \quad
\beta^{\tau}_{V \pi} = \frac{\tau_{q \pi}}{\tau_{V}} \rightarrow 0 \, , \quad
\beta^{l}_{W\pi} = \frac{l_{q\pi}}{\tau_{W}} \rightarrow \frac{1}{5} \, ,\quad
\beta^{l}_{V\pi} = \frac{l_{q\pi}}{\tau_{V}} \rightarrow - h\frac{\beta_0}{20} \, ,
\end{eqnarray}
and
\begin{eqnarray}
\beta_\pi &\rightarrow& \frac{3}{4p_0}\, , \quad 
\beta^{\lambda}_{\pi W} = \frac{\lambda_{\pi W}}{\tau_{\pi}} \rightarrow \frac{n_0}{12 \beta_0 p_0} \, , \quad
\beta^{\lambda}_{\pi V} = \frac{\lambda_{\pi V}}{\tau_{\pi}} \rightarrow -\frac{n_0}{3 \beta^2_0 p_0} \, , \quad
\delta_2 = \frac{4}{3}\beta_{\pi} \, ,\\ 
\beta^{\tau}_{\pi W} &=& \frac{\tau_{\pi W}}{\tau_{\pi}} \rightarrow -\frac{5}{3} \, , \quad
\beta^{\tau}_{\pi V} = \frac{\tau_{\pi V}}{\tau_{\pi}} \rightarrow \frac{8}{3 \beta_0} \, , \quad
\beta^{l}_{\pi W} = \frac{l_{\pi W}}{\tau_{\pi}} \rightarrow \frac{1}{3} \, ,\quad
\beta^{l}_{\pi V} = \frac{l_{\pi V}}{\tau_{\pi}} \rightarrow -\frac{2}{3\beta_0} \, . 
\end{eqnarray}
%

%%%
\section{Conclusions}

In this work we derived the equations of second order dissipative fluid dynamics 
with all first- and second-order terms from the Boltzmann equation using 
the 14-moment method with a linearized collision integral.
We also expressed all coefficients multiplying the second order terms 
independent of the choice of frame and showed that some of these coefficients are different 
in different frames.

%%%
\begin{acknowledgement}

The authors thank T.S.\ Bir\'o, L.P.\ Csernai, P.\ Danielewicz, and P.\ V\'an for valuable remarks.

B.B.\ acknowledges support by
the Alexander von Humboldt foundation via the Feodor Lynen fellowship. 
E.M.\ acknowledges support by OTKA/NKTH 81655, the Computational Subatomic Physics project
(171247/V30) at the University of Bergen, and
the Alexander von Humboldt foundation. 
The work of H.N.\ was supported by
the Extreme Matter Institute (EMMI).

This work was supported by the Helmholtz International Center
for FAIR within the framework of the LOEWE program 
launched by the State of Hesse.

\end{acknowledgement}

%%%

\end{document}